\documentclass[twocolumn]{aastex631}

\newcommand\kms{km~s$^{-1}$}

\newcommand\FeII{\hbox{{\rm Fe~}\kern 0.1em{\sc ii}}}
\newcommand\HeIlam{\hbox{{\rm He~}\kern 0.1em{\sc i}~$\lambda 5877$}}
\newcommand\HeI{\hbox{{\rm He~}\kern 0.1em{\sc i}}}
\newcommand\HeIIlamopt{\hbox{{\rm He~}\kern 0.1em{\sc ii}~$\lambda 4687$}}
\newcommand\HeII{\hbox{{\rm He~}\kern 0.1em{\sc ii}}}
\newcommand\CIV{\hbox{{\rm C~}\kern 0.1em{\sc iv}}}
\newcommand\Hbetalam{\hbox{{\rm H}\kern 0.1em{\sc $\beta$}~$\lambda 4861$}}
\newcommand\Hbeta{\hbox{{\rm H}\kern 0.1em{\sc $\beta$}}}
\newcommand\OIIIlam{\hbox{[{\rm O~}\kern 0.1em{\sc iii}]~$\lambda\lambda 4960, 5008$}}
\newcommand\OIIIlamweak{\hbox{[{\rm O~}\kern 0.1em{\sc iii}]~$\lambda 4960$}}
\newcommand\OIIIlamstrong{\hbox{[{\rm O~}\kern 0.1em{\sc iii}]~$\lambda 5008$}}
\newcommand\OIII{\hbox{[{\rm O~}\kern 0.1em{\sc iii}]}}
\newcommand\Halpha{\hbox{{\rm H}\kern 0.1em{\sc $\alpha$}}}

\accepted{March 14, 2023}
\submitjournal{PASP}

\shorttitle{Revisit NLS1 Methods}
\shortauthors{Khatu et al.}

\begin{document}

\title{Revisiting Emission-Line Measurement Methods for Narrow-Line Active Galactic Nuclei}

\author[0000-0002-0581-6506]{Viraja C. Khatu}
\affiliation{Department of Physics and Astronomy \& Institute of Earth and Space Exploration, The University of Western Ontario, 1151 Richmond Street, London, Ontario N6A 3K7, Canada}
\email{vkhatu@uwo.ca}

\author[0000-0001-6217-8101]{Sarah C. Gallagher}
\affiliation{Department of Physics and Astronomy \& Institute of Earth and Space Exploration, The University of Western Ontario, 1151 Richmond Street, London, Ontario N6A 3K7, Canada}

\author[0000-0003-1728-0304]{Keith Horne}
\affiliation{School of Physics and Astronomy, University of St Andrews, North Haugh, St Andrews, KY16 9SS, Scotland, UK}

\author[0000-0002-8294-9281]{Edward M. Cackett}
\affiliation{Department of Physics \& Astronomy, Wayne State University, 666 W Hancock Street, Detroit, Michigan 48201, USA}

\author{Chen Hu}
\affiliation{Key Laboratory for Particle Astrophysics, Institute of High Energy Physics, Chinese Academy of Sciences, 19B Yuquan Road, Beijing 100049, People's Republic of China}

\author[0000-0002-5830-3544]{Pu Du}
\affiliation{Key Laboratory for Particle Astrophysics, Institute of High Energy Physics, Chinese Academy of Sciences, 19B Yuquan Road, Beijing 100049, People's Republic of China}

\author[0000-0001-9449-9268]{Jian-Min Wang}
\affiliation{Key Laboratory for Particle Astrophysics, Institute of High Energy Physics, Chinese Academy of Sciences, 19B Yuquan Road, Beijing 100049, People's Republic of China}

\author[0000-0002-2121-8960]{Wei-Hao Bian}
\affiliation{Physics Department, Nanjing Normal University, Nanjing 210097, People's Republic of China}

\author{Jin-Ming Bai}
\affiliation{Yunnan Observatories, The Chinese Academy of Sciences, Kunming 650011, People's Republic of China}

\author{Yong-Jie Chen}
\affiliation{Key Laboratory for Particle Astrophysics, Institute of High Energy Physics, Chinese Academy of Sciences, 19B Yuquan Road, Beijing 100049, People's Republic of China}

\author{Patrick Hall}
\affiliation{Department of Physics and Astronomy, York University, 4700 Keele Street, Toronto, Ontario M3J 1P3, Canada}

\author{Bo-Wei Jiang}
\affiliation{Key Laboratory for Particle Astrophysics, Institute of High Energy Physics, Chinese Academy of Sciences, 19B Yuquan Road, Beijing 100049, People's Republic of China}

\author[0000-0003-3823-3419]{Sha-Sha Li}
\affiliation{Yunnan Observatories, The Chinese Academy of Sciences, Kunming 650011, People's Republic of China}

\author[0000-0001-5841-9179]{Yan-Rong Li}
\affiliation{Key Laboratory for Particle Astrophysics, Institute of High Energy Physics, Chinese Academy of Sciences, 19B Yuquan Road, Beijing 100049, People's Republic of China}

\author[0000-0002-9416-8001]{Sofia Pasquini}
\affiliation{Department of Physics and Astronomy, The University of Western Ontario, 1151 Richmond Street, London, Ontario N6A 3K7, Canada}

\author{Yu-Yang Songsheng}
\affiliation{Key Laboratory for Particle Astrophysics, Institute of High Energy Physics, Chinese Academy of Sciences, 19B Yuquan Road, Beijing 100049, People's Republic of China}

\author{Chan Wang}
\affiliation{Physics Department, Nanjing Normal University, Nanjing 210097, People's Republic of China}

\author{Ming Xiao}
\affiliation{Key Laboratory for Particle Astrophysics, Institute of High Energy Physics, Chinese Academy of Sciences, 19B Yuquan Road, Beijing 100049, People's Republic of China}

\author{Zhe Yu}
\affiliation{Key Laboratory for Particle Astrophysics, Institute of High Energy Physics, Chinese Academy of Sciences, 19B Yuquan Road, Beijing 100049, People's Republic of China}

\begin{abstract}

Measuring broad emission-line widths in active galactic nuclei (AGN) is not straightforward owing to the complex nature of flux variability in these systems.  Line-width measurements become especially challenging when signal-to-noise is low, profiles are narrower, or spectral resolution is low.  We conducted an extensive correlation analysis between emission-line measurements from the optical spectra of Markarian~142 (Mrk~142; a narrow-line Seyfert galaxy) taken with the Gemini North Telescope (Gemini) at a spectral resolution of $185.6\pm10.2$~\kms\ and the Lijiang Telescope (LJT) at $695.2\pm3.9$~\kms\ to investigate the disparities in the measured broad-line widths from both telescope data.  Mrk~142 posed a challenge due to its narrow broad-line profiles, which were severely affected by instrumental broadening in the lower-resolution LJT spectra.  We discovered that allowing the narrow-line flux of permitted lines having broad and narrow components to vary during spectral fitting caused a leak in the narrow-line flux to the broad component, resulting in broader broad-line widths in the LJT spectra.  Fixing the narrow-line flux ratios constrained the flux leak and yielded the \Hbeta\ broad-line widths from LJT spectra $\sim$54\% closer to the Gemini \Hbeta\ widths than with flexible narrow-line ratios.  The availability of spectra at different resolutions presented this unique opportunity to inspect how spectral resolution affected emission-line profiles in our data and adopt a unique method to accurately measure broad-line widths.  Reconsidering line-measurement methods while studying diverse AGN populations is critical for the success of future reverberation-mapping studies.  Based on the technique used in this work, we offer recommendations for measuring line widths in narrow-line AGN.
\newline{{\em Unified Astronomy Thesaurus concepts}: \href{http://astrothesaurus.org/uat/16}{Active galactic nuclei (16)}; \href{http://astrothesaurus.org/uat/1558}{Spectroscopy (1558)}}
\newline{{\em Supporting material}: Machine-readable tables and spectra}

\end{abstract}

\section{Introduction} \label{sec:intro}

Observing gaps in multi-epoch astronomical data are commonplace.  Sparsely sampled ground-based observations are mainly a result of weather conditions and pressure from other programs (for queue observatories).  Data gaps can impact the programs requiring frequent visits to the sky, e.g., time-series observations.  One way to work around data gaps is by performing simultaneous observations with different telescopes.  However, using data from various observatories together in a meaningful way can be challenging because of the need for sufficient cross-calibration.

Combining spectroscopic observations from different facilities requires careful consideration of various factors (e.g., exposure time, seeing, spectral resolution, etc.) affecting the data.  When employing multiple telescopes to observe the same spectral region for a given data set, the instrument specifications used with different facilities may not be identical.  As an example, if the data from two telescopes were taken with different slit widths, their spectral resolutions may not match.  The wider the slit, the lower the spectral resolution (depending on the grating used).  Instrumental broadening in lower-resolution spectra causes line features to appear broader than the true value or sometimes blended with neighboring lines.  This broadening effect can result in inaccurate measurements of the physical parameters, e.g., full width at half maximum (FWHM) and flux of an emission line.  However, the severity of the effect may be different for different studies.  For objects showing broad-line profiles considerably narrower than the typical AGN population, e.g., Narrow-Line Seyfert~1 galaxies (NLS1s) accreting at super-Eddington rates, accounting for instrumental broadening effects while performing line measurements becomes more critical.

Historically, reverberation-mapping \citep[RM;][]{blandford_mckee_1982, peterson1993} studies have focused mainly on low-redshift AGN ($z<0.3$) that typically follow sub-Eddington accretion \citep[e.g.,][]{stirpe_etal_1994, santos-lleo_etal_1997, collier_etal_1998, dietrich_etal_1998, kaspi_etal_2000, bentz_etal_2006} instead of the more atypical super-Eddington objects (accreting well above the Eddington limit).  RM is a powerful technique that takes advantage of the variability in AGN over a range of timescales \citep[from several days to weeks and years; e.g,][]{peterson_etal_1982} -- broad-line region (BLR) response on larger scales to the continuum variations from the accretion disk, with a positive time lag -- and provides a way to convert the time lag into a spatial distance, the size of the BLR.  The BLR size along with the width of an emission line is used to obtain black hole masses in AGN.  Applying the RM technique successfully to measure AGN black hole masses thus requires accurate measurements of the emission-line widths.  A majority of the low-redshift AGN exhibit broad emission lines (e.g., \Hbeta, \Halpha, \CIV, etc.) with $\text{FWHM}\geq 2000$~\kms\ in their spectra.  Recent efforts by the Super-Eddington Accreting Massive Black Holes (SEAMBH) collaboration \citep{du_etal_2014, wang_etal_2014, hu_etal_2015, du_etal_2015, du_etal_2016_1, du_etal_2016_2, du_etal_2018} have successfully identified $\sim$20 highly accreting AGN to date.  SEAMBH objects show spectral features containing narrower \Hbeta\ broad lines (FWHM$_{\mathrm{\Hbeta}} \lesssim 2000$~\kms), weak \OIII, and strong optical \FeII\ emission lines that appear as a bumpy pseudo-continuum, similar to other NLS1s \citep{osterbrock_pogge_1987, boroson_green_1992, boller_etal_1996, veroncetty_etal_2001}.  As we branch out to studying different categories of AGN, and given the importance of accurate line measurements for RM analysis, we must carefully reconsider: (1) {\em How does spectral resolution influence the shapes of emission lines in our data?}; and (2) {\em Are the current methods of measuring emission-line properties sufficient or are they limited in any way to achieving our desired science goals?}

The aim of this paper is to address the above two questions with the optical spectra of the super-Eddington AGN, Markarian~142 (Mrk~142), taken with the Gemini North Observatory (Gemini) and the Lijiang Telescope (LJT).  As a part of the broader RM campaign, \citet{cackett_etal_2020} performed overlapping photometric and spectroscopic observations of Mrk~142 with telescope facilities worldwide.  The goal of the campaign is to study the structure of the accretion disk and the BLR simultaneously, for the first time, in a super-Eddington AGN.  Because the spectral observations of Mrk~142 with Gemini suffered with gaps due to unfavourable weather, the Gemini observations were complemented with simultaneous observations from LJT.  With the Gemini+LJT data, an ultraviolet (UV) continuum to \Hbeta\ lag for the object was measured for the first time \citep{khatu_etal_2022}.  The Gemini and LJT spectra were taken with the same grating but different slit widths -- 0.75\arcsec\ for Gemini and 2.5\arcsec\ for LJT; therefore, their spectral resolutions (corresponding to the instrumental FWHMs) differed -- $185.6\pm10.2$~\kms\ for Gemini and $695.2\pm3.9$~\kms\ for LJT.  Here, the challenge was to accurately measure the emission lines for spectra at considerably different spectral resolutions for a narrow-line object.  In this paper, we discuss how to address this challenge with our spectral-fitting procedure.  Further, we provide recommendations of the best strategies for performing accurate emission-line measurements in narrow-line AGN.

This paper is organized as follows.  Section~\ref{sec:methods} details the process of preparing the data for correlation analysis and the correlations between the Gemini and LJT spectral measurements.  Section~\ref{sec:results} outlines the results.  We discuss the results in Section~\ref{sec:discussion} along with our recommendations.  Section~\ref{sec:conclusion} summarizes our findings.

\section{Methodology} \label{sec:methods}

We pre-processed the Gemini and LJT calibrated spectra independently through {\tt\string PrepSpec}\footnote{{\tt\string PrepSpec} is a spectral analysis software that fits the continuum and emission lines in the input spectra with a composite model through an iterative process while correcting for any relative deviations in the calibrated wavelength and flux scales of the spectra.  See Section~\ref{subsec:prepspec_preprocessing} for more details.} (developer: K. Horne) to correct for any relative deviations from their calibrations.  We then modelled the continuum and emission lines in the spectra with {\tt\string Sherpa}\footnote{{\tt\string Sherpa} is a software application for modeling and fitting astronomical images and spectra.  In this work, the {\tt\string Sherpa} v4.10.0 application was used within Coronagraphic Imager with Adaptive Optics ({\tt\string CIAO}) v4.10.0, the X-ray Data Analysis Software designed by the Chandra X-ray Center.  For full documentation of {\tt\string CIAO}-{\tt\string Sherpa}, see \url{https://cxc.harvard.edu/sherpa4.14/}.} \citep{freeman_etal_2001} v4.10.0 with a {\tt\string Python} wrapper script to examine the correlations between the Gemini and LJT spectral measurements.

\subsection{{\tt\string PrepSpec} Pre-processing} \label{subsec:prepspec_preprocessing}

{\tt\string PrepSpec} corrects for differences in relative wavelength and flux calibrations of the spectra by modeling their continuum, emission lines, and absorption lines.  However, {\tt\string PrepSpec} requires that spectra have no gaps in wavelength and extremely large flux values, for example, resulting from cosmic ray hits.  To prepare the Gemini spectra for {\tt\string PrepSpec}, we applied the following corrections: (1) replaced detector gaps, and residual features from cosmic-ray correction and sky subtraction with interpolated and simulated data; (2) recovered the flat spectral regions with zero flux values (resulting from the position angle of the slit non-parallel to the parallactic angle) and bumpy regions (resulting from the calibration process) by using a high signal-to-noise (S/N) mean spectrum as the reference spectrum; and (3) recovered slit losses (resulting from the 0.75\arcsec\ narrow slit used for observations) using 5.00\arcsec\ wide-slit spectra taken on the same nights.  We developed multiple scripts in {\tt\string Python} v3.6.5 to perform the above corrections.  All {\tt\string Python} scripts are publicly available in the {\tt\string GitHub} repository, {\em prepdataps} (\url{https://github.com/Virachan/prepdataps}).  Information on the usage and citation of the scripts is provided in the repository.  The LJT spectra had no gaps and hence were directly processed through {\tt\string PrepSpec} after calibration.  The {\tt\string GitHub} repository includes the user manual for {\tt\string PrepSpec} including a detailed tutorial with the Mrk~142 Gemini data set.

\subsection{Gemini/LJT Correlation Analysis} \label{subsec:glcorrelation}

\begin{figure*}[ht!]
\epsscale{1.15}
\plotone{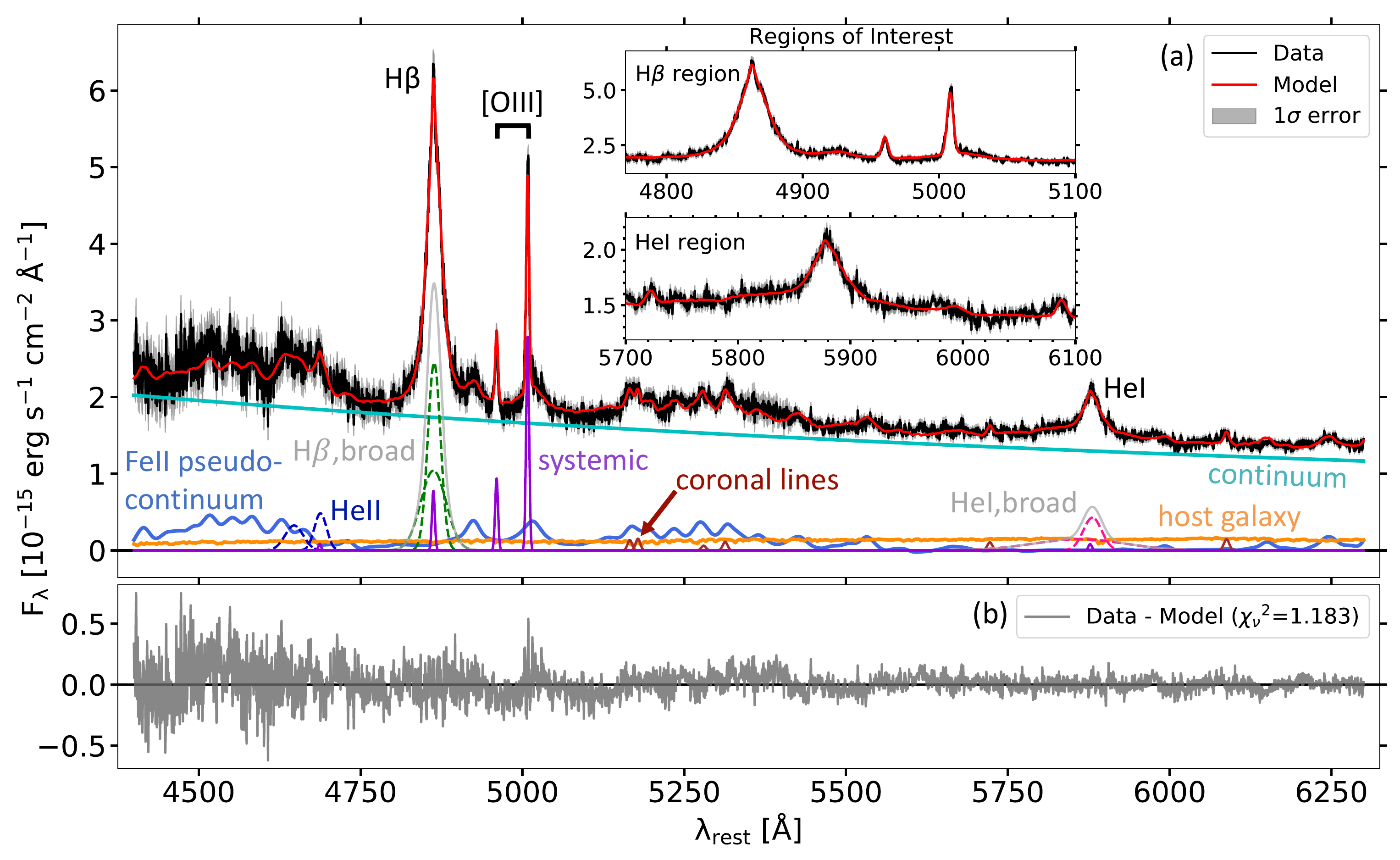}
\caption{Composite model fit to epoch 24 of the Mrk~142 Gemini data displaying individual components of the model.  Panel {\em a}: Composite model (red solid curve) fit to the data (black solid curve) from 4430~\AA\ to 6300~\AA\ is shown in the main panel, and the \Hbeta\ and \HeI\ {\em Regions of Interest} are shown in the {\em Inset} panels.  The individual components of the model are displayed at the bottom of the panel: continuum (cyan solid curve); \FeII\ I~Zw~1 template as a pseudo-continuum (faint blue solid curve); host-galaxy template (orange solid curve); \HeII\ broad components (blue dashed Gaussians); \Hbeta\ broad components (green dashed Gaussians); \HeI\ broad components (pink dashed Gaussians); narrow-line components of \Hbeta, \HeI, and \OIII\ (purple solid Gaussians); and high-ionization coronal lines (brown solid Gaussians).  The total broad \Hbeta\ and \HeI\ profiles are also overplotted (gray solid curves).  Panel {\em b}: Residuals of the model with reduced $\chi^2$, ${\chi_{\nu}}^2=1.183$.  The model shows larger residuals around the \OIIIlamstrong\ line indicating a sub-optimal fit in that region.  The noisier blue end of the spectrum affects the overall fit in that region, thus resulting in larger residuals compared to the red end of the spectrum.
\label{fig:spectral_model_single_epoch_gemini}}
\end{figure*}

\begin{figure*}
\epsscale{1.15}
\plotone{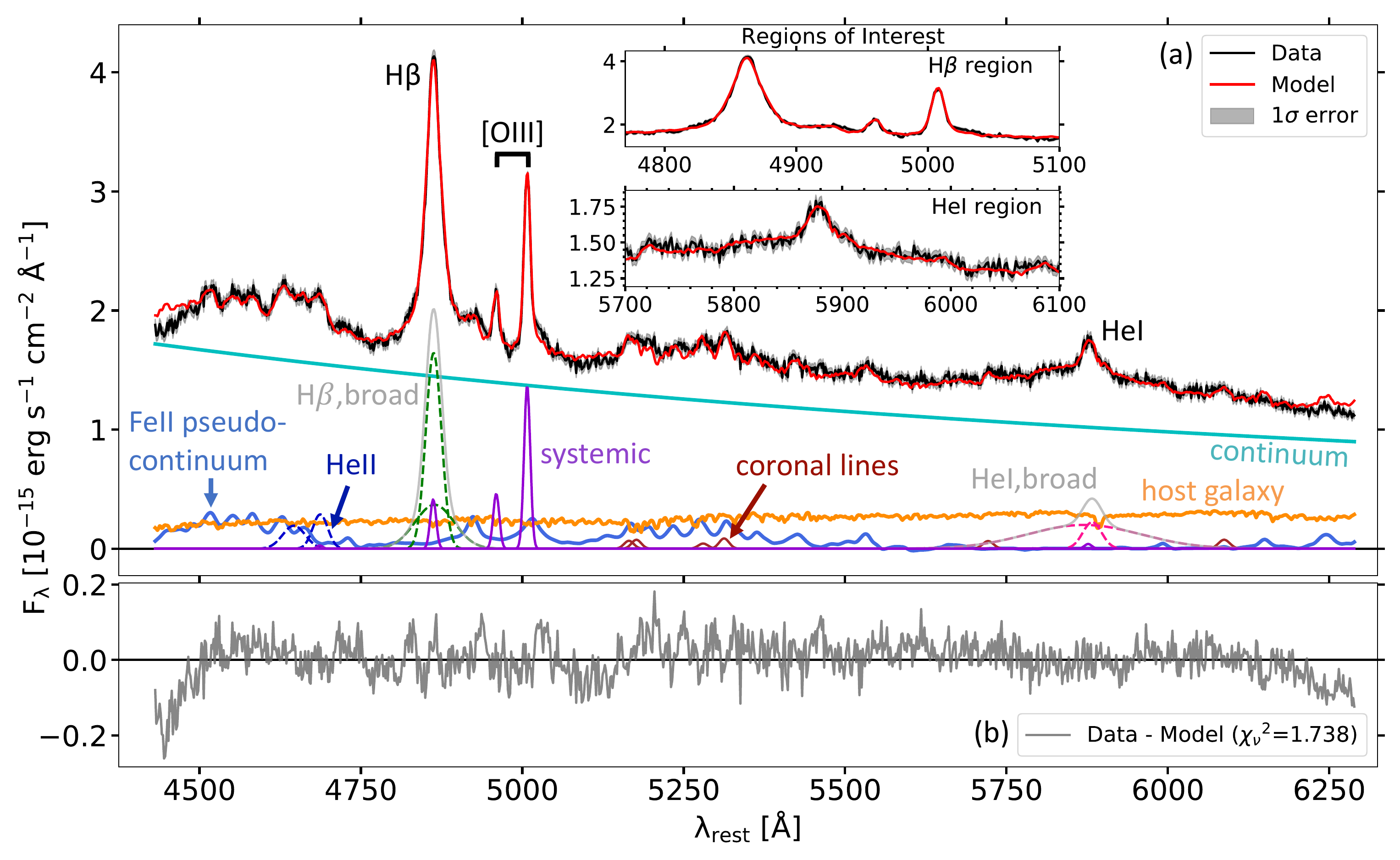}
\caption{Composite model fit to epoch 24 of the Mrk~142 LJT data displaying individual components of the model.  See caption of Figure~\ref{fig:spectral_model_single_epoch_gemini} for a description of the individual model components in Panel {\em a}.  The red side of the broad \Hbeta\ emission line shows contamination with the \FeII\ emission at $\sim$4923~\AA.  Similarly, the \OIIIlamstrong\ line shows considerable blending with the \FeII\ feature in its red wing, thus affecting a reliable measurement of the \OIIIlamstrong\ line.  Panel {\em b} shows the residuals of the model with ${\chi_{\nu}}^2=1.711$.  The smaller residuals indicate an overall good fit to the spectrum.  The model performance drops significantly at both ends of the spectrum although it does not impact measurements in the {\em Regions of Interest}.
\label{fig:spectral_model_single_epoch_ljt}}
\end{figure*}

We investigated correlations between the Mrk~142 Gemini and LJT emission-line measurements to study the effect of low spectral resolution on the measured FWHM values.  To fit the \Hbetalam\ and \HeIlam\ {\em Regions of Interest}, we initially followed a standard approach.  We set the FWHM, flux, and position of the \OIIIlamstrong\ line to vary during the fit, and fixed the FWHMs and positions of the narrow components in \Hbeta\ and \HeI\ relative to the \OIIIlamstrong\ line.  We kept the \Hbeta\ and \HeI\ narrow-line fluxes as free parameters.  Further, we used two Gaussians to model the broad \Hbeta\ and \HeI\ emission lines.  We fixed the FWHM of the broader broad \Hbeta\ component and allowed the width of the narrower broad component to be a free parameter to reduce the number of free parameters and avoid degeneracy in the spectral model.  For the \HeI\ line, we fixed the FWHMs of both its broad components with respect to the FWHM of the flexible, narrower broad \Hbeta\ component.  We applied the \citet{boroson_green_1992} template to fit the \FeII\ pseudo-continuum and the \citet{bruzual_charlot_2003} host galaxy template with 11~Gyr at $z=0.05$ to correct for the effect of host-galaxy emission.  This approach worked well for the Gemini spectra.  However, the LJT \Hbeta\ FWHM measurements were broader than expected, indicating that the fitting approach failed for the LJT spectra.  Figures~\ref{fig:spectral_model_single_epoch_gemini} and \ref{fig:spectral_model_single_epoch_ljt} show model fits to a single-epoch spectrum from Gemini and LJT, respectively.

Due to low resolution of the LJT spectrum (695.2~\kms), emission-line features appear smeared as compared to the Gemini spectrum (185.6~\kms).  In \Hbeta, the sharper peak of the narrow component evident in the Gemini spectrum is not clearly distinguishable in the LJT spectrum.  The \FeII\ emission-line component on the red side of \Hbeta\ also appears to contaminate the broader \Hbeta\ component.  Similarly, the red wing of the \OIIIlamstrong\ line appears to be blended with the \FeII\ emission on its red side.  The possible contamination with \FeII\ emission due to low spectral resolution of the LJT spectra results in a broad FWHM for the \OIIIlamstrong\ line.  Further, at the instrumental resolution of the LJT spectrograph, the \OIII\ lines with FWHM~$\sim 320$~\kms\ (as measured from the higher-resolution Gemini spectra) are unresolved.  The unresolved \OIII\ lines in the LJT spectra further complicate the measurements of their widths.

\subsubsection{The Problem -- Posed: {\it Why do the broad \Hbeta\ FWHMs from the Gemini and LJT spectra not match?}} \label{subsubsec:problem}

\begin{figure}[ht!]
\epsscale{1.15}
\plotone{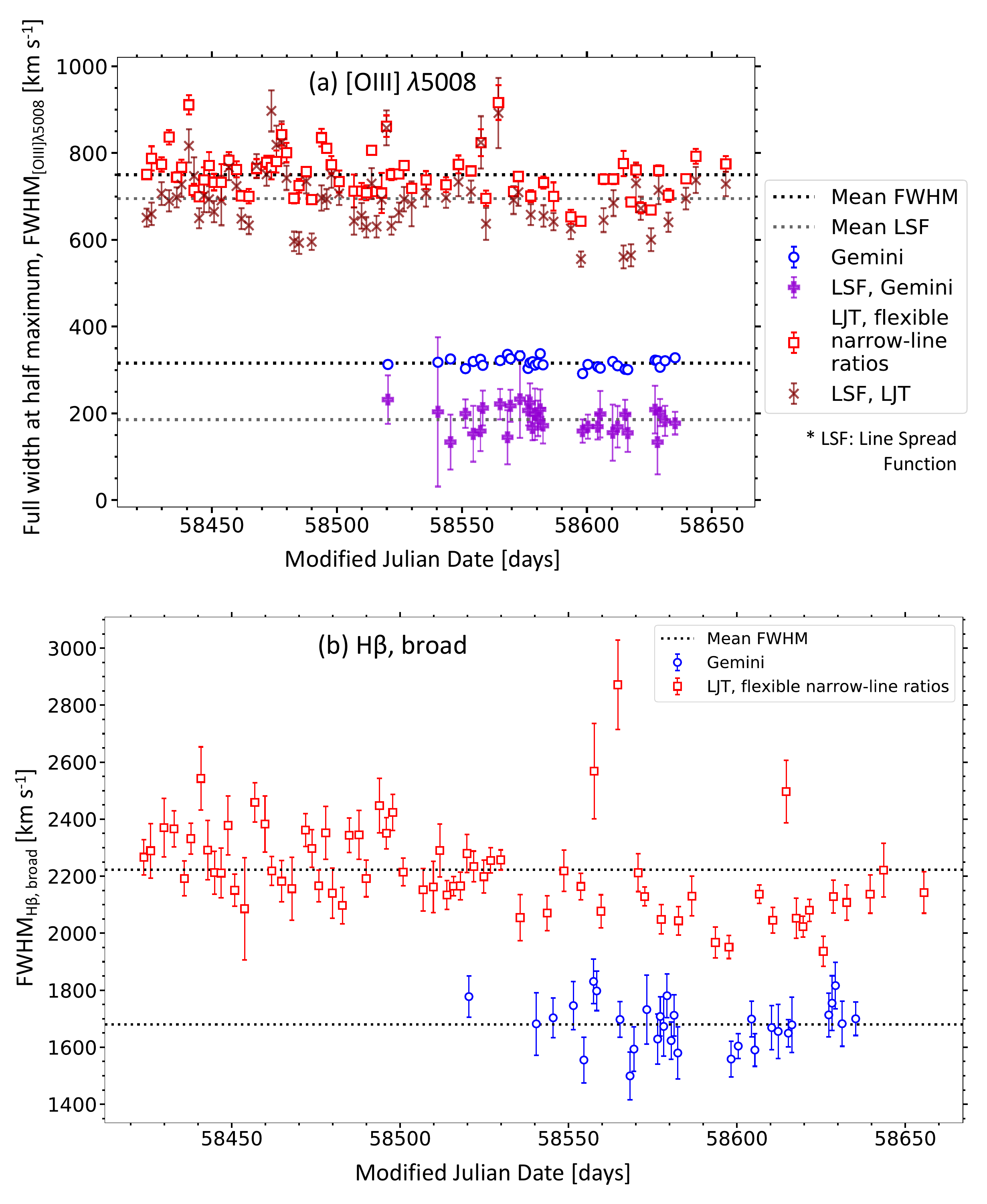}
\caption{Differences in the full width at half maximum (FWHM) values measured from the Gemini (blue open circles) and LJT (red open circles) spectra for \OIIIlamstrong\ (panel {\em a}) and broad \Hbeta\ (panel {\em b}), indicating that the LJT measurements have a broader effective width of 670.0~\kms\ (see text for the definition) as expected owing to the Gemini and LJT instrumental resolutions.  In panel {\em a}, the purple pluses and maroon crosses represent the line spread functions (determined from the comparison star observed in the same slit as the target at each epoch) for the Gemini and LJT data, respectively. 
 The fitting procedure for both the Gemini and the LJT spectra incorporated flexible narrow-line ratio of \Hbeta\ to \OIIIlamstrong.
\label{fig:fwhm_5008_hbeta_b}}
\end{figure}

We found that the FWHM values of the broad \Hbeta\ and the \OIIIlamstrong\ lines in the Gemini and LJT spectra did not agree with each other.  Figure~\ref{fig:fwhm_5008_hbeta_b} displays the differences in the measured FWHMs of the two lines.  To understand the cause of this problem, it is important to consider the differing instrumental resolutions of Gemini and LJT.  We used the spectra of a G-type comparison star, common for both the Gemini and the LJT data, observed in the same slit as the target for every exposure to calculate the instrumental resolutions.  The comparison star, similar to the target, appears point-like in the slit and does not fill the entire slit unlike arc lamp spectra or sky lines; therefore, unresolved absorption lines in the stellar spectra can be used to estimate the line spread function of the instrument more accurately than arc lamp spectra or sky lines.  We fit the weak and hence unresolved \Hbeta\ stellar absorption line close to the \Hbeta/\OIII\ complex in the Gemini data using a Gaussian function, where the FWHM of the Gaussian gave the mean resolution of $185.6\pm10.2$~\kms\ with reference to the \OIIIlamstrong\ line.  Following the procedure in \citet{du_etal_2016_1}, we convolved a higher-resolution stellar template spectrum with a Gaussian function to obtain the observed stellar spectra in the LJT data; here as well, the FWHM of the convolved Gaussian provided the mean resolution of $695.2\pm3.9$~\kms\ for the LJT spectrograph.  Considering the difference in the resolutions of the Gemini and LJT spectra, we expect the FWHM values from LJT to have 670.0~\kms\ broader effective width (here, defined as the width of the Gaussian kernel required to smooth the higher-resolution Gemini spectrum to the lower-resolution LJT spectrum, and given by $\sqrt{695.2^2-185.6^2}$ for Gaussian profiles) than the Gemini FWHM values.  The \OIIIlamstrong\ line in the LJT spectra (Figure~\ref{fig:fwhm_5008_hbeta_b}, panel {\em a}) has $\sim$680~\kms\ broader effective widths than the \OIIIlamstrong\ FWHMs measured from the Gemini spectra as expected from the differing instrumental resolutions of the two telescopes.  However, the LJT \Hbeta\ broad line shows a much greater effect, $\sim$1450~\kms\ broader effective widths than those measured from the Gemini spectra.  The broad \HeI\ FWHM measurements were affected in the same way as \Hbeta.  This implies that there are multiple factors influencing the FWHM measurements in the LJT spectra, likely resulting from instrumental broadening.

\begin{figure}[ht!]
\epsscale{1.15}
\plotone{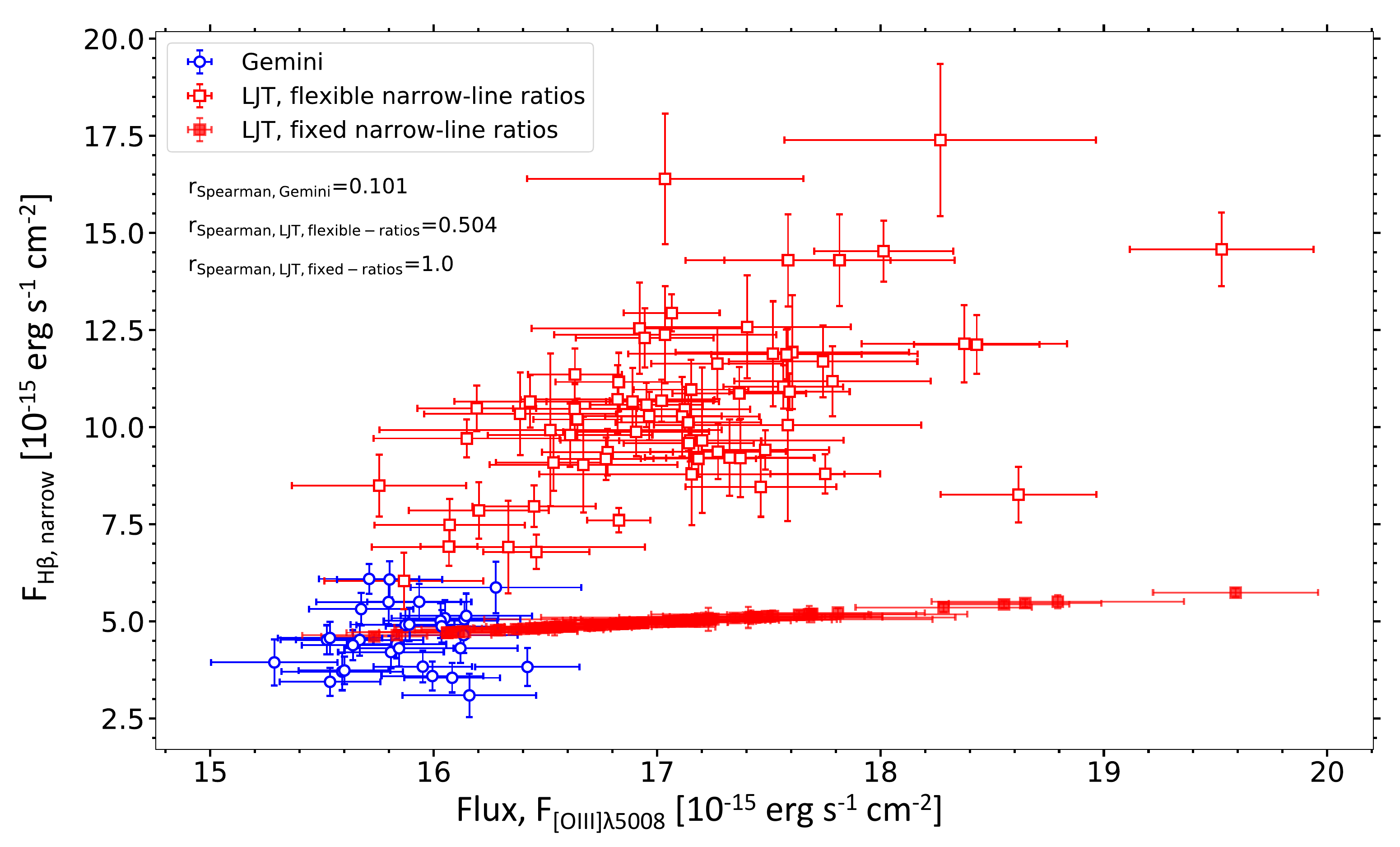}
\caption{Correlations between the narrow \Hbeta\ line flux and the \OIIIlamstrong\ line flux from the Gemini (blue circles) and LJT (red squares) spectra.  Open symbols represent measurements with flexible narrow-line flux ratios of $F_{\mathrm{\Hbeta}}/F_{\mathrm{\OIIIlamstrong}}$ and $F_{\mathrm{\HeI}}/F_{\mathrm{\OIIIlamstrong}}$, whereas filled symbols (for LJT only) represent measurements with fixed narrow-line flux ratios.  For fitting the LJT spectra, we fixed the above narrow-line flux ratios determined from the Gemini spectra fitting process (0.293 for \Hbeta\ to \OIIIlamstrong\ and 0.034 for \HeI\ to \OIIIlamstrong) as the resolved \OIIIlamstrong\ line in the Gemini spectra provided reliable measurements of the narrow-line flux ratios.  The weak positive correlation for the LJT measurements with flexible narrow-line ratios indicates that the ``broader'' \OIIIlamstrong\ line results in an excess narrow-line flux for the narrow \Hbeta\ component.  
\label{fig:fixed_flux_hbeta_to_5008}}
\end{figure}

To identify the possible reason(s) for the observed discrepancies in the FWHM measurements, we considered the narrow-line fluxes in the \Hbeta\ and \OIIIlamstrong\ lines that we allowed to vary during spectral fitting.  Figure~\ref{fig:fixed_flux_hbeta_to_5008} displays the correlation between the measured fluxes for the two lines.  For the LJT spectra, the fluxes in the narrow \Hbeta\ line (red open squares) are correlated to the \OIIIlamstrong\ line fluxes ($r_{\mathrm{Spearman}}\sim0.5$) with a large scatter in the measured values from both lines.  However, the fluxes for narrow lines, originating in the narrow-line region (NLR) of AGN, are not expected to vary relative to the continuum variations over the timescale of an RM campaign.  The constancy of the narrow-line fluxes is attributed to the NLR being farther away from the central source (supermassive black hole + accretion disk) than the BLR, which receives the continuum variations.  The correlation observed between the LJT narrow \Hbeta\ and \OIIIlamstrong\ line fluxes is absent in the Gemini flux values (blue open circles; $r_{\mathrm{Spearman}}\sim0.1$), which are also more constrained than the LJT measurements.

\subsubsection{The Culprit -- Pronounced: {\it A ``broader''} {\rm \OIIIlamstrong} {\it FWHM leads to a broader broad \Hbeta\ FWHM in the LJT spectra.}} \label{subsubsec:culprit}

Integrating the observations from Figures~\ref{fig:fwhm_5008_hbeta_b} and \ref{fig:fixed_flux_hbeta_to_5008}, we concluded that the ``broader'' (than expected) \OIIIlamstrong\ line in the LJT spectra causes the correlation and scatter observed in the \Hbeta\ narrow-line fluxes, consequently leading to the ``broader'' broad \Hbeta\ FWHMs in the LJT spectra as compared to the narrower broad \Hbeta\ in the Gemini spectra.  The lower spectral resolution of the LJT spectra smears the narrow-line profile of the \OIIIlamstrong\ line causing it to appear broader than usual (see Figure~\ref{fig:spectral_model_single_epoch_ljt}).  This effect is amplified in the case of Mrk~142 as it is a NLS1 with narrower and weaker \OIII\ lines compared to the more typical broad-line objects (with narrow-line widths of several hundred \kms).  Also, the likely contamination from the \FeII\ emission at the red wing of \OIIIlamstrong\ (see Figure~\ref{fig:spectral_model_single_epoch_ljt}) further complicates the picture.  The net result is a ``broader'' FWHM measured for the \OIII\ line.  We interpret this result to mean the following: \\
{\em Because the width of the narrow \Hbeta\ component is set equal to width of the} \OIII\ {\em line, a ``broader''} \OIII\ {\em leads to a ``broader'' narrow \Hbeta\ in the LJT spectra.  The ``broader'' narrow \Hbeta\ collects more flux leaving less flux for the broad \Hbeta\ components, which are kept flexible in the model.  The more the flux in the narrow \Hbeta\ line, the lower the peak flux of the broad \Hbeta\ (after subtracting the narrow component), resulting into a ``broader'' broad \Hbeta\ FWHM.}

\subsubsection{The Solution -- Proposed: {\it Tie the flux in narrow \Hbeta\ relative to the flux in} {\rm \OIIIlamstrong}.} \label{subsubsec:solution}

To contain the effect of the ``broader'' \OIIIlamstrong\ line on the broad \Hbeta\ FWHM measurements in the LJT spectra, we fixed the narrow-line fluxes as well as the widths of \Hbeta\ and \HeI\ relative to the \OIII\ flux from the Gemini spectral measurements.  At the resolution of the Gemini spectra (185.6~\kms), the \OIIIlamstrong\ line is just resolved.  Thus, the FWHM measurements for the \OIII\ line in the Gemini spectra are more reliable than the LJT values.  We therefore applied the narrow-line flux ratios, $F_{\mathrm{\Hbeta}}/F_{\mathrm{\OIIIlamstrong}}=0.293$ and $F_{\mathrm{\HeI}}/F_{\mathrm{\OIIIlamstrong}}=0.034$, from the Gemini spectra to the \Hbeta\ and \HeI\ lines, respectively, in the LJT spectra.

Tying the LJT narrow-line flux ratios confined the flux measured for the \Hbeta\ narrow component as compared to the \Hbeta\ narrow-line flux measured when the narrow-line flux ratios were kept flexible.  Figure~\ref{fig:fixed_flux_hbeta_to_5008} displays the fixed \Hbeta\ narrow-line fluxes from the LJT spectra (red filled squares) as exactly correlated ($r_{\mathrm{Spearman}}=1.0$) to the LJT \OIIIlamstrong\ fluxes.  The large scatter in the narrow \Hbeta\ line fluxes from flexible line ratios for the LJT spectra (red open squares; $\sigma^2=4.631$) is considerably reduced after adopting fixed narrow-line ratios ($\sigma^2=0.042$).  This reduction in scatter signifies that the \Hbeta\ narrow-line fluxes in the LJT spectra are now well constrained. \\
{\em The applied flux ratios along with fixed widths for the narrow lines helped constrain the narrow \Hbeta\ flux.  With a fixed width and constrained flux of the narrow \Hbeta\ component, the ``excess'' flux is directed back to the broad components.  The more the broad \Hbeta\ flux, the higher the peak flux measured for the broad \Hbeta, resulting into a narrower broad \Hbeta\ FWHM.}

\section{Results} \label{sec:results}

We studied the effect of the LJT spectral resolution (695.2~\kms) on the measured \Hbeta\ FWHMs by correlating the emission-line measurements in the Mrk~142 Gemini and LJT spectra.  We identified that the ``broader'' FWHM measurements of the unresolved \OIIIlamstrong\ line in the LJT spectra lead to an overestimation of the \Hbeta\ narrow-line fluxes by a factor of two, ultimately resulting in \Hbeta\ FWHM values with $\sim$1450~\kms\ broader effective width than the \Hbeta\ FWHM values from the Gemini spectra.  To correct for this effect, we adapted our spectral fitting procedure for the LJT spectra by fixing the narrow-line ratios of \Hbeta\ and \HeI\ with respect to the \OIIIlamstrong\ line based on our spectral measurements for the Gemini spectra containing resolved \OIII\ lines.  We present our results below.

\begin{figure}[ht!]
\epsscale{1.15}
\plotone{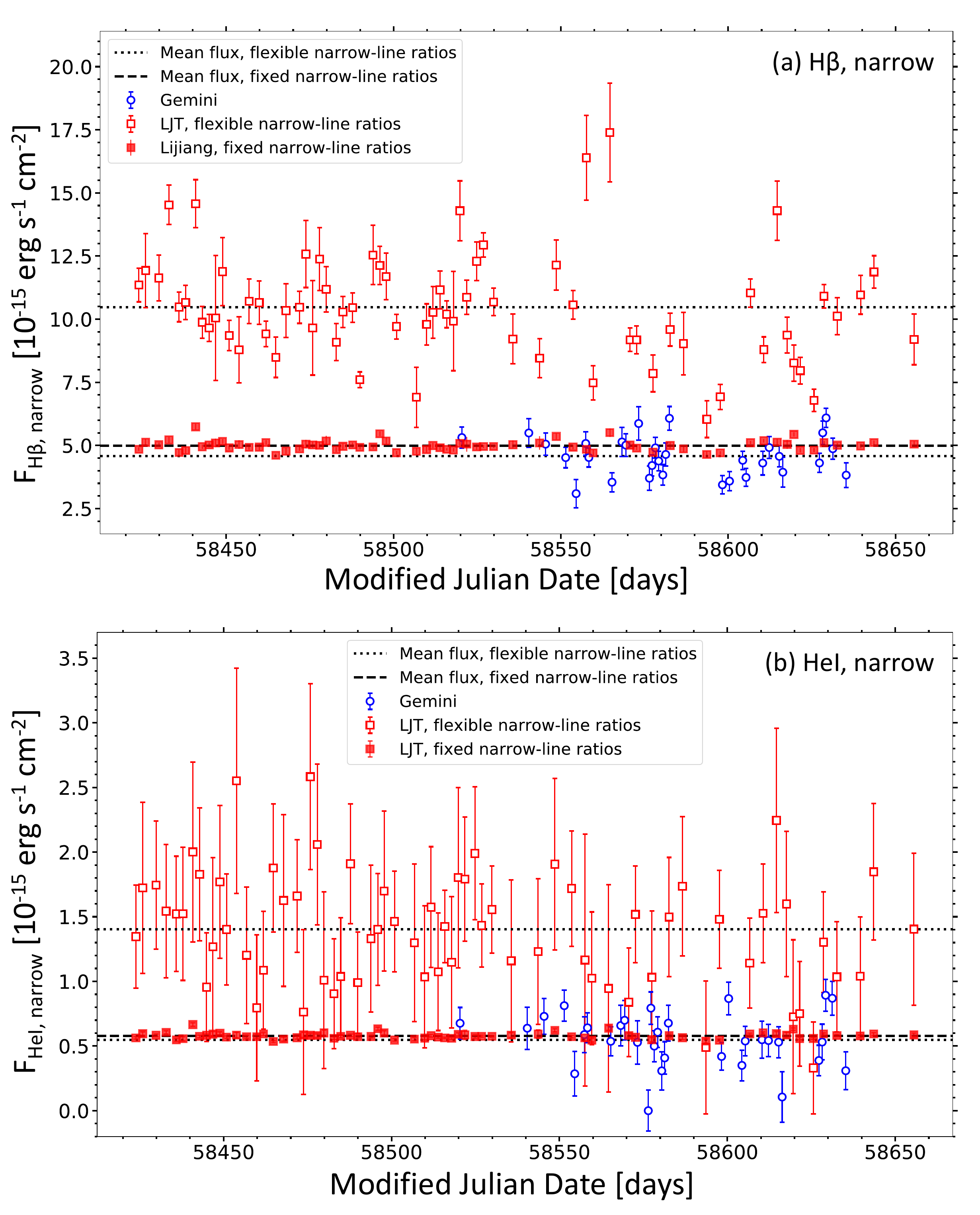}
\caption{Narrow \Hbeta\ (panel {\em a}) and \HeI\ (panel {\em b}) light curves measured from the Gemini (blue circles) and LJT (red squares) spectra fit with flexible (open symbols) and fixed (filled symbols) narrow-line flux ratios.  Fixing the \Hbeta\ and \HeI\ narrow-line flux ratios of 0.293 and 0.034, respectively, relative to the \OIIIlamstrong\ line flux considerably reduced the scatter in the LJT measurements as well as resulted in lower uncertainties.  With fixed narrow-line flux ratios, the new fluxes for the narrow \Hbeta\ and \HeI\ lines from the LJT spectra are closer to the Gemini measurements.
\label{fig:lightcurves_hbeta_hei_after_fix}}
\end{figure}

Applying narrow-line flux ratios for fitting the lower-resolution LJT spectra significantly reduces the scatter in the narrow-line fluxes of the \Hbeta\ and \HeI\ broad lines.  Figure~\ref{fig:lightcurves_hbeta_hei_after_fix} displays the fluxes of the narrow component in \Hbeta\ (panel {\em a}) and \HeI\ (panel {\em b}) before and after applying the flux ratios for the LJT spectra.  Before fixing the narrow-line fluxes, the higher \Hbeta\ to \OIIIlamstrong\ and \HeI\ to \OIIIlamstrong\ flux ratios for the lower-resolution LJT spectra result from the ``excess'' flux gathered under the narrow \Hbeta\ and \HeI\ lines owing to the narrow \Hbeta\ and \HeI\ line widths set equal to the width of the ``broader'' \OIIIlamstrong\ line.  After fixing the narrow-line flux ratios, the \Hbeta\ and \HeI\ line measurements are more consistent, with values similar to the narrow-line fluxes from the Gemini data.  The higher resolution of the Gemini spectra, providing resolved \OIII\ lines, allowed precise measurements of \OIII\ line widths and fluxes, thus resulting in the smaller scatter in the \Hbeta\ and \HeI\ narrow-line fluxes.  In addition, we expect the narrow-line fluxes to be constant over the length of our campaign.  However, the \Hbeta\ and \HeI\ narrow-line fluxes from the LJT spectra are unreliable when the narrow-line ratios are free parameters due to the large uncertainties and scatter in the measured values.  The constant \Hbeta\ and \HeI\ narrow-line flux is clearly evident after fixing the narrow-line ratios for the LJT spectra.

\begin{figure}[ht!]
\epsscale{1.15}
\plotone{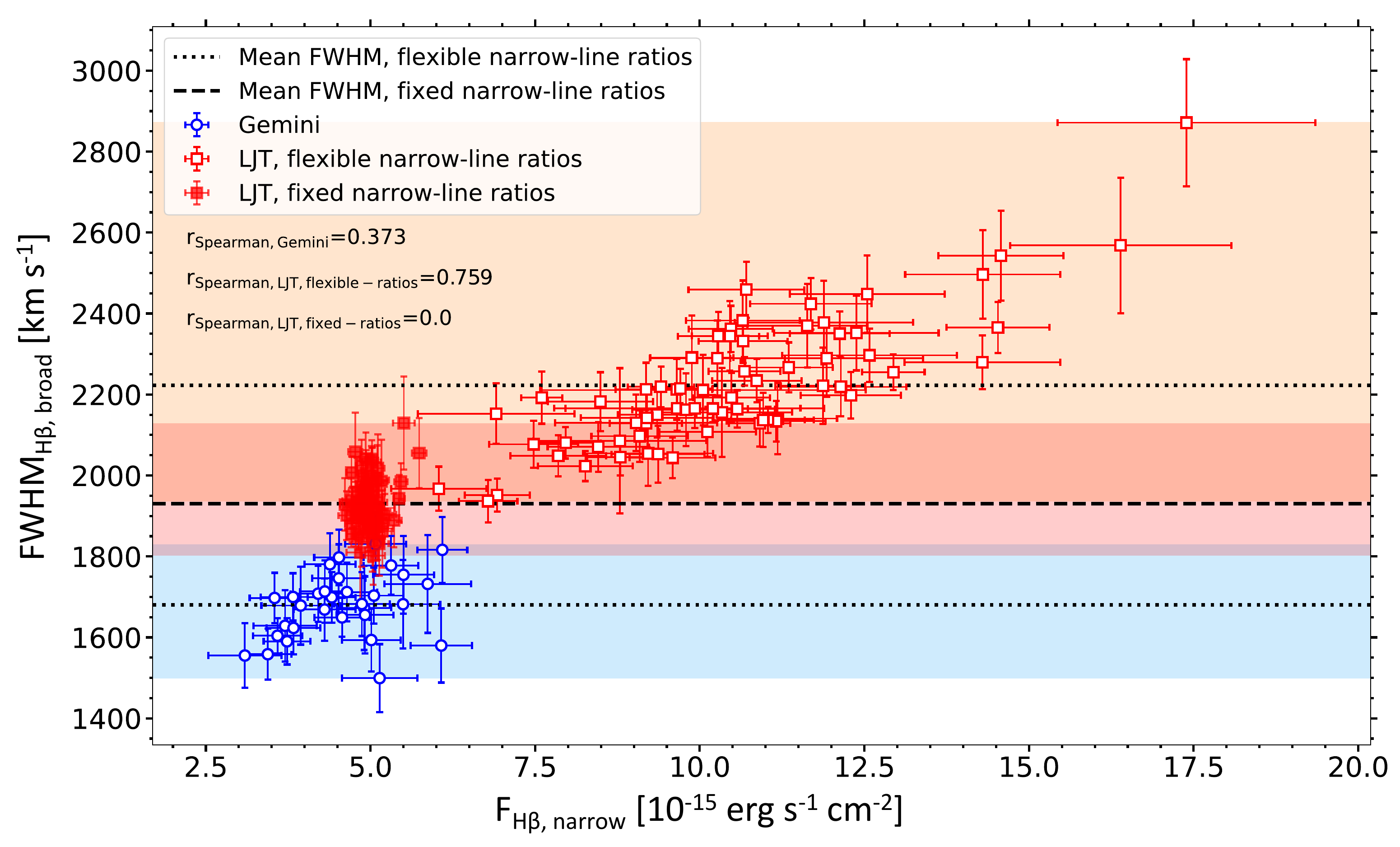}
\caption{Correlations between the broad \Hbeta\ full width at half maximum (FWHM) and the narrow \Hbeta\ flux from the Gemini (blue circles) and LJT (red squares) spectra fit with flexible (open symbols) and fixed (filled symbols) narrow-line flux ratios.  The scatter in the LJT measurements with flexible narrow-line ratios (orange shaded region) is considerably reduced (red shaded region) after fixing the narrow-line ratios.  Fixed ratios nullified the strong correlation that is evident in the LJT measurements with flexible ratios.
\label{fig:hbeta_fwhm_flux_dependence}}
\end{figure}

\begin{figure}[ht!]
\epsscale{1.15}
\plotone{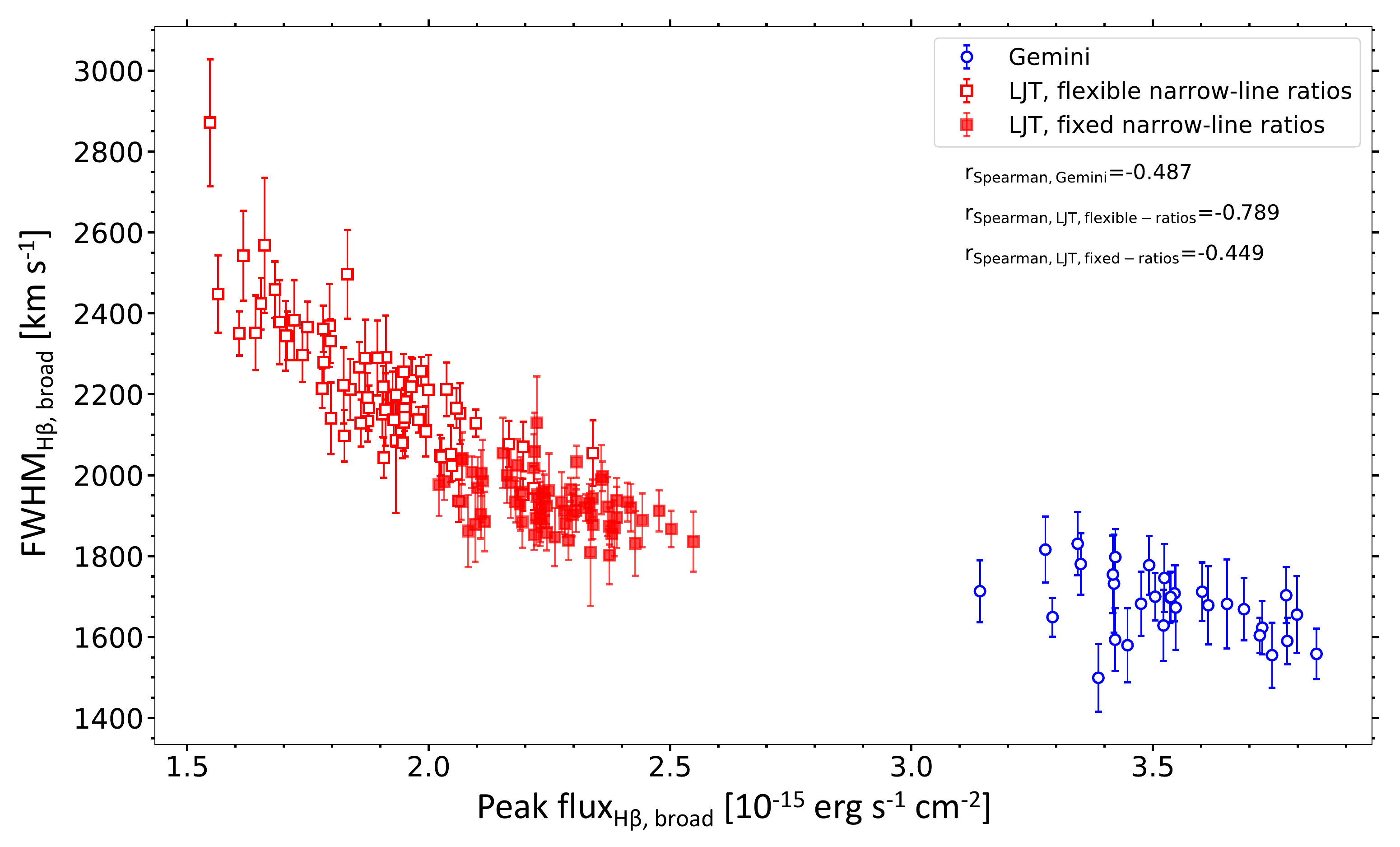}
\caption{Dependence of the full width at half maximum (FWHM) on the peak flux of the broad \Hbeta\ line from the Gemini (blue circles) and LJT (red squares) spectra fit with flexible (open symbols) and fixed (filled symbols) narrow-line flux ratios.  Fixed ratios nullified the strong correlation that is evident in the LJT measurements with flexible ratios.
\label{fig:fwhm_hbeta_peakflux_dependence}}
\end{figure}

We found that the constrained \Hbeta\ narrow-line fluxes in the LJT spectra yield narrower FWHMs for the \Hbeta\ broad component, and Figure~\ref{fig:hbeta_fwhm_flux_dependence} displays this result.  Flexible narrow-line flux ratios in the LJT spectra result in a large scatter in the broad \Hbeta\ FWHMs (red open squares) with a mean value of $\sim$2220~\kms.  After fixing the narrow-line flux ratios, the constrained narrow-line flux in \Hbeta\ re-allocates the ``excess'' flux to the broad components.  As a result of the higher broad-line peak flux, the half-maximum point increases, thus measuring a narrower \Hbeta\ broad-line profile.  Figure~\ref{fig:fwhm_hbeta_peakflux_dependence} explicitly shows the dependence of the measured \Hbeta\ FWHM on the peak flux of the line.  With flexible narrow-line ratios for the lower-resolution LJT spectra, the measured broad \Hbeta\ FWHMs are strongly negatively correlated ($r_{\mathrm{Spearman}}=-0.789$) with the peak fluxes of the line.  This correlation is weaker ($r_{\mathrm{Spearman}}=-0.449$) with the fixed narrow-line flux ratios for the LJT spectra and also similar to the correlation observed for the Gemini spectra ($r_{\mathrm{Spearman}}=-0.487$).  The remaining dependence of the broad \Hbeta\ FWHM on its peak flux can be attributed to the contamination in the \Hbeta\ line from the nearby \FeII\ emission (also discussed below).  The new measurements of the broad \Hbeta\ FWHM, obtained after fixing the narrow-line flux ratios for the LJT spectra, have a mean of $\sim$1930~\kms, $\sim$54\% closer to the Gemini \Hbeta\ mean FWHM.  The new broad \Hbeta\ FWHM values also show a scatter of $\sim$340~\kms, similar to the scatter in the Gemini \Hbeta\ FWHM values (see Figure~\ref{fig:hbeta_fwhm_flux_dependence}).

\begin{figure}[ht!]
\epsscale{1.15}
\plotone{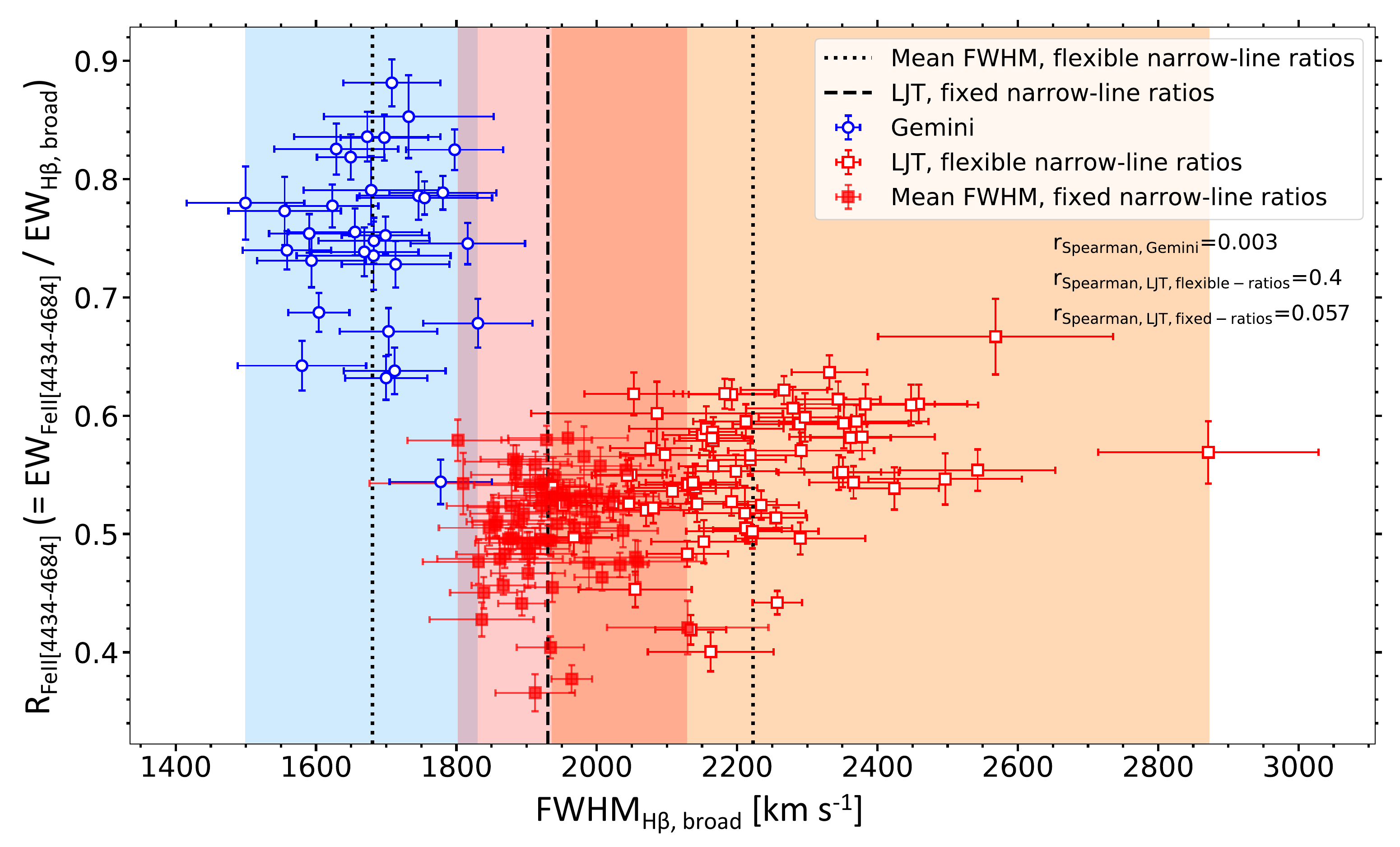}
\caption{Correlations between the strength of \FeII\ (ratio of equivalent width of \FeII\ in the [4434--4684~\AA] region to that of broad \Hbeta) and the full width at half maximum (FWHM) of the broad \Hbeta\ from the Gemini (blue circles) and LJT (red squares) spectra fit with flexible (open symbols) and fixed (filled symbols) narrow-line flux ratios.  Fixing the narrow-line flux ratios moved the mean \Hbeta\ FWHM from the LJT data closer to the mean of the Gemini \Hbeta\ FWHM measurements.  There is an unexplained offset between the Gemini and the new LJT measurements likely due to the \FeII\ emission contaminating the broad \Hbeta\ line in the LJT spectra.
\label{fig:feii_hbeta_fwhm_dependence}}
\end{figure}

We also examined the correlation between the strength of \FeII\ ($R_{\mathrm{\FeII}}$, defined as the ratio of equivalent width of \FeII\ in the 4434--4684~\AA\ region to that of broad \Hbeta) and the FWHM of broad \Hbeta, shown in Figure~\ref{fig:feii_hbeta_fwhm_dependence}.  The $R_{\mathrm{\FeII}}$--FWHM$_{\mathrm{\Hbeta,broad}}$ correlation reiterates the issue of the broader \Hbeta\ FWHM measurements in a different context.  The LJT \Hbeta\ FWHM values with flexible narrow-line ratios (red open squares) are weakly correlated to $R_{\mathrm{\FeII}}$.  This correlation vanishes after fixing the narrow-line flux ratio of \Hbeta\ to \OIIIlamstrong.  The new broad \Hbeta\ measurements (red filled squares) are closer to the values measured from the Gemini spectra; however, an offset is still present.  At the low spectral resolution of the LJT spectra, the \FeII\ emission around the \Hbeta\ line likely contaminates the \Hbeta\ broad-line emission (see Figure~\ref{fig:spectral_model_single_epoch_ljt}).  The remainder offset noted above can likely be a result of the potential \FeII\ contamination with the broad \Hbeta\ line in the LJT spectra.  Therefore, understanding any interplay between the two lines is crucial.  In addition, we note that the $R_{\mathrm{\FeII}}$ measurements from the LJT spectra are smaller as compared to those from the Gemini spectra, likely due to the lower resolution of the LJT data.

\section{Discussion} \label{sec:discussion}

We correlated emission-line measurements from Mrk~142 spectra taken with Gemini and LJT at different resolutions (185.6~\kms\ for the former and 695.2~\kms\ for the latter) to study the discrepancies observed in the measured FWHM values of broad \Hbeta.  Given the NLS1 nature of Mrk~142, its broad \Hbeta\ profile is narrower as compared to the more typical broad-line Seyfert~1s, and therefore challenging to accurately measure in lower-resolution spectra.  Accurate measurements of the narrow-line fluxes are critical to measure broad emission-line widths, which in turn are used to derive the black hole masses in AGN.

Our adopted method of measuring the broad \Hbeta\ FWHM decreased the FWHM values measured from the LJT spectra significantly, thus bringing them closer to the Gemini measurements as well as reducing the scatter in the measured values.  The smaller scatter in the \Hbeta\ FWHMs from the LJT spectra was an improvement by a factor of $\sim$2.6 as compared to the measurements with flexible narrow-line ratios.  Although we significantly improved our measurements from the LJT spectra, the new \Hbeta\ FWHM values do not exactly match with those from the Gemini spectra.  There is an offset of $\sim$250~\kms\ between the mean FWHM values from the two data sets that is perhaps a consequence of the \FeII\ emission contaminating the broad, red wing of \Hbeta.  At lower resolution, the strong \FeII\ lines in the LJT spectra are smeared as compared to the sharper features in the Gemini spectra (see Figures~\ref{fig:spectral_model_single_epoch_gemini} and \ref{fig:spectral_model_single_epoch_ljt}), resulting in weaker \FeII\ measured from the LJT spectra (see Figure~\ref{fig:feii_hbeta_fwhm_dependence}).  Therefore, there likely exists cross-talk between the two broad lines affecting the measurements of both the \FeII\ emission and the broad \Hbeta\ wings.  

\begin{figure}[ht!]
\epsscale{1.15}
\plotone{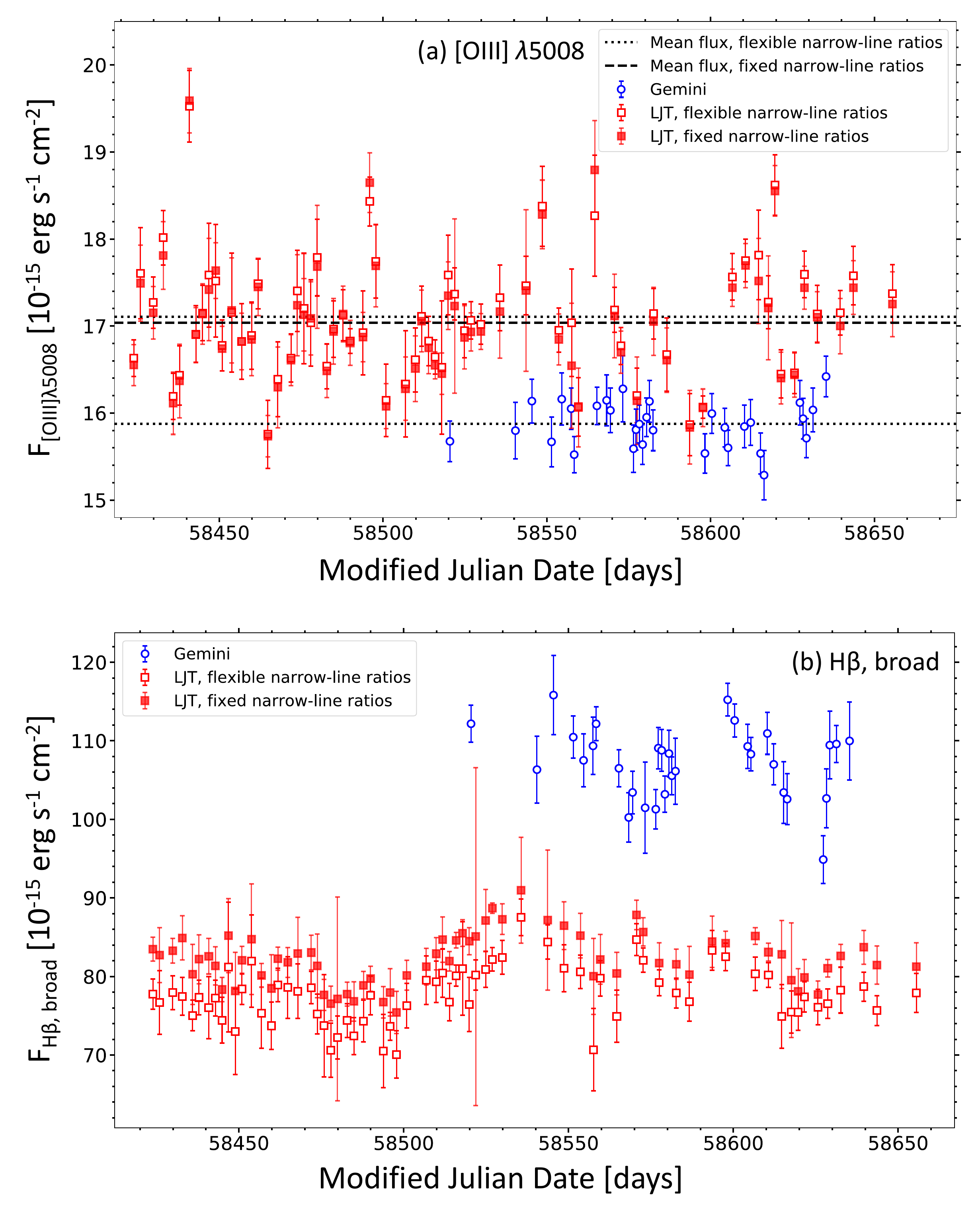}
\caption{\OIIIlamstrong\ (panel {\em a}) and broad \Hbeta\ (panel {\em b}) light curves from the Gemini (blue circles) and LJT (red squares) spectra fit with flexible (open symbols) and fixed (filled symbols) narrow-line flux ratios.  Because the \OIIIlamstrong\ line flux is kept flexible even while fitting with fixed narrow-line flux ratios, the measured fluxes before and after fixing narrow-line ratios are similar for the \OIII\ line.  However, there is $\sim$10\% increase in the the broad \Hbeta\ flux from the LJT spectra fit with fixed narrow-line ratios.
\label{fig:lightcurves_hbeta_5008_after_fix}}
\end{figure}

We also caution that our method does not affect the \OIII\ line measurements from the LJT spectra although it restores the broad-line flux measured for the \Hbeta\ line.  Figure~\ref{fig:lightcurves_hbeta_5008_after_fix} makes this explicit.  The \OIII\ fluxes measured from the LJT spectra are approximately the same (red filled squares) before fixing the narrow-line flux ratios (see in Figure~\ref{fig:lightcurves_hbeta_5008_after_fix}, panel {\em a}).  The \OIII\ fluxes remain unchanged because the \OIIIlamstrong\ flux was freed during spectral fitting.  In contrast, an increase in flux is clearly seen in the broad \Hbeta\ light curve (red filled squares in Figure~\ref{fig:lightcurves_hbeta_5008_after_fix}, panel {\em b}).  Fixing the narrow-line flux ratios resulted in an increase of $\lesssim$10\% in the \Hbeta\ broad-line flux in the LJT spectra -- a difference that is greater than or similar to the \Hbeta\ flux uncertainties in a majority of the LJT epochs.  Our proposed method here can be used with spectral data sets having different resolutions from either different gratings or different slit widths.  However, it does not test the effect on line fluxes, e.g., the offset observed in the broad \Hbeta\ line fluxes from Gemini and LJT (see Figure~\ref{fig:lightcurves_hbeta_5008_after_fix}, panel {\em b}).  Such an offset may result from varying slit widths or orientation of the slit or a combined effect of both.  The orientation effect can be especially concerning for extended sources in which the \OIII\ emission emerging from the NLR on larger spatial scales and the broad \Hbeta\ emission from the BLR closer to the central black hole can fill the slit differently.  However, the size of the central source in Mrk~142 is $3\arcsec \times 3\arcsec$ (and the apparent size of the host galaxy is $15\arcsec \times 15\arcsec$), where the issue of \Hbeta/\OIII\ emission filling the slit differently on a spatial scale is likely negligible.  Nevertheless, we note that testing our proposed method with different slit widths and orientations will help understand the effects on measured line widths and fluxes in greater detail.

In general, the root-mean-square (RMS) spectra do not suffer from the blending of broad- and narrow-line features because the variable part of the spectrum is exclusively from the BLR.  However, RMS spectra are noisier as compared to the spectra from individual epochs; the emission-line widths obtained from RMS spectra have higher uncertainties.  Therefore, we considered measuring \Hbeta\ broad-line widths from individual-epoch Gemini and LJT spectra.

\subsection{Comparison to Previous Studies} \label{subsec:previous_comparison}

We compared our methodology to the techniques used in previous studies.  In a previous SEAMBH monitoring campaign with LJT \citep{du_etal_2014, hu_etal_2015}, Mrk~142 spectra were taken with the same instrument settings as used for the data set in this work.  \citet{hu_etal_2015} describe the spectral fitting technique adopted for measuring \Hbeta\ FWHM from the previous campaign.  They followed a two-step process: (1) Assuming no narrow-line contribution in the \Hbeta\ emission and given that the narrow lines do not vary at the short variability timescales of the broad-line gas, they modeled the \Hbeta\ profile with a Gauss-Hermite function.  (2) They modeled the \Hbeta\ narrow line with a Gaussian whose width was set equal to that of the \OIIIlamstrong\ line and its position fixed relative to the \OIII\ line.  They ran their fits in two cycles -- first, assuming 10\% of the \OIIIlamstrong\ flux for the narrow \Hbeta\ component, and the value of the broad \Hbeta\ FWHM was used from this run; and second, assuming 20\% narrow-line flux, and the uncertainty on the FWHM measurement was used from this run in addition to the uncertainty from the Gauss-Hermite fit for the line.  While the calibration of the uncertainties on the measured \Hbeta\ FWHMs from the two-cycle fits and Gauss-Hermite fits appears sufficient, the flux ratio of the \Hbeta\ to \OIII\ narrow lines is somewhat arbitrarily defined.  As the narrow-line flux ratios can be different for different objects, using arbitrary ratios or even the same ratios for different objects can result in a systematic overestimation or underestimation of the broad \Hbeta\ FWHM values.  With the opportunity of using Mrk~142 spectra at two different resolutions, we applied the more reliable narrow-line flux ratios determined from the higher-resolution and high S/N Gemini spectra to the lower-resolution LJT spectra for the \Hbeta\ FWHM measurements from the latter.

Two of the proposed methods of correcting for the effect of different spectral resolution on line-width measurements include: (1) fitting a difference spectrum between an input spectrum, whose flux scale factor (or flux ratio) is to be determined, and a reference spectrum, typically a high S/N spectrum \citep{vanGroningen_wanders_1992}; and (2) subtracting the RMS width of the line spread function of the spectrograph from the observed line profile \citep{fausnaugh_etal_2017}.  In the method by \citet{vanGroningen_wanders_1992}, if the input and reference (at higher resolution) spectra have different resolutions, the algorithm first finds a convolution factor to degrade the reference spectrum and then determines the scale factor by fitting a simple, analytical function to the difference spectrum that is non-variable over the spectral regions of narrow lines.  However, the success rate of determining the scale factor for the correct input spectrum decreases as the difference in the spectral resolution decreases.  For the latter case, the method suggested in this work provides a robust way of measuring the line width by first calculating the flux scale factor from a higher-resolution spectrum and then applying it to lower-resolution spectra without confusing between spectra with less dissimilar resolutions.  An alternative method by \citep{fausnaugh_etal_2017} attempts to first determine the line spread function of the spectrograph to subtract from an observed profile so that the intrinsic line width is measured more accurately, especially for low-resolution spectra.  However, determination of a spectrograph's line spread function relies on a prior measurement of a narrow-line profile for at least one of the objects in the data set \citep[e.g., the \citealt{whittle1992} database used by][]{fausnaugh_etal_2017}.  Furthermore, using a previous database value to correct for spectrograph's line spread function does not necessarily catch the temporal variations in the instrument behavior, where narrow-line ratios from high-resolution spectra taken on the same night of the low-resolution spectra, as in this work, are beneficial for accurate measurement of the line width.

Additionally, past studies discuss the use of different line-width measures for reliable black hole mass estimates \citep{liu_bian_2022}.  \citet{peterson_etal_2004} conducted a time-series analysis of 35 AGN to test the effectiveness of different line-width measures for calculating black hole masses.  They conclude that the line dispersion ($\sigma_{\mathrm{line}}$; or the RMS width calculated from the second moment of a line profile) is a more robust measure of the variable line profile than FWHM, especially in objects with strong narrow lines.  However, \citet{bian_etal_2008} argue that $\sigma_{\mathrm{line}}$ strongly depends on the contribution from the wings of broad lines.  They analyzed 329 NLS1s from the Sloan Digital Sky Survey catalog to determine their black hole masses using $\sigma_{\mathrm{line}}$.  They infer that in the cases where the \Hbeta\ broad profile is defined by two broad components (as for Mrk~142 spectra in this work), the black hole mass measurements using $\sigma_{\mathrm{line}}$ are about 0.5~dex larger than those obtained from FWHM measurements.  

In another study, \citet{collin_etal_2006} characterized the broad \Hbeta\ emission-line profiles of all reverberation-mapped AGN to then based on the ratio of $\mathrm{FWHM}/\sigma_\mathrm{line}$$^{\mathrm[}$\footnote{The nature of a broad-line profile determines the relationship between the FWHM and $\sigma_{\mathrm{line}}$, where $\mathrm{FWHM}/\sigma_\mathrm{line}=2.35$ for a Gaussian profile.}$^{\mathrm]}$, separating the AGN with narrower broad \Hbeta\ profiles ($\mathrm{FWHM}/\sigma_\mathrm{line}<2.35$) from those with broader \Hbeta\ ($\mathrm{FWHM}/\sigma_\mathrm{line}>2.35$).  In their analysis of the virial product ($VP=M_{\bullet}/f$, where f is scale factor) with FWHM and $\sigma_{\mathrm{line}}$ from both mean and RMS spectra, they showed: (1) although $\sigma_{\mathrm{line}}$ yields consistent results for scale factors from both the mean and the RMS spectra for objects with narrower as well as broader broad \Hbeta\ profiles, $\sigma_{\mathrm{line}}$ from the mean spectrum is on average $\sim$20\% broader in the mean than the RMS spectra; and (2) for the narrower broad \Hbeta\ population (including NLS1s), although the scale factors from FWHM are larger by a factor of $\sim$3 from both mean and RMS spectra, FWHM is only $\sim$10\% broader in the mean than the RMS spectra.  It is worth noting from the above discussions that both the FWHM and the $\sigma_{\mathrm{line}}$ are typically measured to be broader in the mean than the RMS spectra.

\subsection{Recommendations for Line-Width Measurements in Narrow-Line AGN} \label{subsec:recommendations}

The discrepancies highlighted by previous studies emphasize that accurately measuring broad-line widths in narrow-line objects, e.g., NLS1s, is not straightforward.  As demonstrated in this work, the flux leak from the broad-line to the narrow-line components can significantly affect the measured FWHM of the broad lines, e.g., \Hbeta, in NLS1s.  The flux leak from \FeII\ around \Hbeta\ can also affect the \Hbeta\ FWHM measurements.  A consistent watch on how the quality of the data and measurement methods influence each other is essential.  \Hbeta, surrounded by \FeII\ emission in the optical and the \OIIIlam\ lines, is challenging to measure; however, it is accessible to most of the ground-based observatories over a considerably wide redshift range and therefore prominently used for RM studies.  Based on our analysis in this work, we provide recommendations on measuring emission-line widths in narrow-line AGN.

{\em We strongly recommend both higher- and lower-resolution observations for RM analysis of narrow-line AGN.}  In RM, the nearest neighboring narrow line is typically used for relative flux calibration of the broad lines of interest, e.g., \OIIIlamstrong\ is used as a calibrator line for measuring \Hbeta.  Therefore, a {\em completely resolved} calibrator line is essential.  We regard this as the primary requirement for accurately measuring the broad-line widths in narrow-line AGN.  Then, while measuring line profiles, applying narrow-line flux ratios (relative to the calibrator line) determined from the higher-resolution spectra will allow more accurate measurements of line widths with the mean spectra.  If only lower-resolution spectra are available, we recommend using higher-resolution spectra from archival observations to determine the appropriate narrow-line flux ratios.  If no archival higher-resolution spectra exist, we highly recommend scheduling {\em at least one} higher-resolution spectrum along with lower-resolution spectra at {\em each epoch}.  RM analysis of narrow-line AGN will benefit extensively with simultaneous higher- and lower-resolution observations.

\section{Conclusion} \label{sec:conclusion}

We performed a detailed correlation analysis of the spectral measurements from the Mrk~142 data taken with Gemini at higher resolution (185.6~\kms) and LJT at lower resolution (695.2~\kms) to understand the effect of different spectral resolutions on the measured physical properties of emission lines.  The FWHMs measured for the broad \Hbeta\ from the Gemini and LJT spectra did not overlap.  Through our analysis, we identified that the ``broader'', unresolved \OIIIlamstrong\ in the LJT spectra affected the \Hbeta\ FWHM measurements during spectral fitting.  We corrected for the LJT \Hbeta\ FWHM values by fixing the narrow-line flux ratio of \Hbeta\ to the \OIIIlamstrong\ line flux as determined from the Gemini spectral fits.  Adopting this procedure, the mean \Hbeta\ FWHM reduced from $\sim$2220~\kms\ to $\sim$1930~\kms, an improvement of $\sim$54\% in the mean value.  We summarize our main results below.

\begin{enumerate}
    \item In Mrk~142, lower-resolution spectra with an unresolved \OIIIlamstrong\ line, a fixed width (equal to the \OIII\ line width) but flexible line flux for the narrow \Hbeta\ caused a flux leak from the broad to the narrow component, resulting in a lower peak flux for the broad-line profile and therefore a broader broad \Hbeta\ FWHM as compared to the higher-resolution spectra with a resolved \OIIIlamstrong\ line.
    \item Fixing the narrow-line flux ratio of \Hbeta\ to \OIIIlamstrong\ while measuring the \Hbeta\ broad-line profile in the LJT spectra nullified the correlation of the broad \Hbeta\ FWHM with the \OIIIlamstrong\ flux as well as reduced the scatter in the FWHM values by a factor of 2.6, equal to the scatter in the \Hbeta\ FWHMs measured from the Gemini spectra.  Consequently, the mean of the \Hbeta\ FWHM values decreased by $\sim$54\% or an effective width of $\sim$1000~\kms.  The remaining offset is likely from \FeII.
    \item Considering the impact of a {\em resolved} \OIIIlamstrong\ on the \Hbeta\ FWHM measurements, we strongly recommend using both higher- and lower-resolution spectra for measuring line profiles in narrow-line AGN.
\end{enumerate}

We leveraged the access to both the Gemini and the LJT spectra of Mrk~142 for RM analysis -- while the LJT observations expanded the time baseline and filled the gaps in Gemini observations, the Gemini spectra at higher resolution allowed more precise \Hbeta\ line measurements from the LJT spectra.  We emphasize that measuring broad-line profiles in objects with narrower broad lines than the typical AGN population is not straightforward.  Branching out to diverse populations for RM studies perhaps needs a revisit to emission-line measurement methods.

\section{acknowledgments} \label{sec:acknowledgements}

V.C.K. thanks T.~A.~Boroson, who provided the \FeII\ template to C.H. for the spectral modeling process.  We acknowledge the support of the Natural Sciences and Engineering Research Council of Canada (NSERC), Discovery Grant RGPIN/04157.  V.C.K. acknowledges the support of the Ontario Graduate Scholarships.  C.H. acknowledges support from the National Science Foundation of China (12122305).  P.D. acknowledges financial support from National Science Foundation of China grants NSFC-12022301, -11873048, and -11991051.  The research of V.C.K. was partially supported by the New Technologies for Canadian Observatories, an NSERC CREATE program.  We acknowledge the support of the National Key R\&D Program of China No. 2021YFA1600404, and the National Science Foundation of China (11991054, 11833008, 11991051, 11973029).  V.C.K. also acknowledges Jonathan Trump, Martin Houde, Stanimir Metchev, and Charles McKenzie for their valuable feedback and discussions.

This research was based on observations obtained at the Gemini Observatory, which is managed by the Association of Universities for Research in Astronomy (AURA) under a cooperative agreement with the National Science Foundation on behalf of the Gemini Observatory partnership: the National Science Foundation (United States), National Research Council (Canada), Agencia Nacional de Investigaci\'{o}n y Desarrollo (Chile), Ministerio de Ciencia, Tecnolog\'{i}a e Innovaci\'{o}n (Argentina), Minist\'{e}rio da Ci\^{e}ncia, Tecnologia, Inova\c{c}\~{o}es e Comunica\c{c}\~{o}es (Brazil), and Korea Astronomy and Space Science Institute (Republic of Korea).  This work was enabled by observations made from the Gemini North telescope, located within the Maunakea Science Reserve and adjacent to the summit of Maunakea.  We are grateful for the privilege of observing the Universe from a place that is unique in both its astronomical quality and its cultural significance.

This research used observations from the Lijiang 2.4-meter Telescope funded by the Chinese Academy of Sciences (CAS) and the People’s Government of Yunnan Province.

\vspace{5mm}

\facilities{Gemini North Telescope (GMOS),
            Lijiang 2.4-meter Telescope (YFOSC)}

\software{{\tt\string Python}, 
          {\tt\string Astropy} \citep{astropy_collaboration_etal_2013}, 
          {\em prepdataps},
          {\tt\string PrepSpec}, 
          {\tt\string Sherpa} \citep{freeman_etal_2001}}

\clearpage

\bibliography{manuscript}{}
\bibliographystyle{aasjournal}

\allauthors

\end{document}